\documentclass[useAMS,usenatbib]{mn2e}

\usepackage{graphicx}
\usepackage{amsmath}
\usepackage{amssymb}
\usepackage{color}
\usepackage{subfig}

\newcommand\msun{\, \rm M_\odot}

\newcommand\kms{\, \rm km\,s^{-1}}

\newcommand\kpc{{\, \rm kpc}}
\newcommand\yr{{\, \rm yr}}

\newcommand\au{{\, \rm AU}}

\newcommand\aout{{a_{\rm out}}}

\title[Secular TDEs by low-mass SMBHs secondaries]{The secular tidal disruption of stars by low-mass Super Massive Black Holes secondaries in galactic nuclei}
\author[G. Fragione]{Giacomo Fragione$^{1}$\thanks{E-mail: giacomo.fragione@mail.huji.ac.il}, Nathan Leigh$^{2,3,4}$\\
$^1$Racah Institute for Physics, The Hebrew University, Jerusalem 91904, Israel\\
$^2$Department of Astrophysics, American Museum of Natural History, New York, NY 10024, USA\\
$^3$Department of Physics and Astronomy, Stony Brook University, Stony Brook, NY 11794-3800, USA\\
$^4$Center for Computational Astrophysics, Flatiron Institute, 162 Fifth Avenue, New York, NY 10010, USA}

\begin{document}

\maketitle

\begin{abstract}
Stars passing too close to a super massive black hole (SMBH) can produce tidal disruption events (TDEs). Since the resulting stellar debris can produce an electromagnetic flare, TDEs are believed to probe the presence of single SMBHs in galactic nuclei, which otherwise remain dark. In this paper, we show how stars orbiting an IMBH secondary are perturbed by an SMBH primary. We find that the evolution of the stellar orbits are severely affected by the primary SMBH due to secular effects and stars orbiting with high inclinations with respect to the SMBH-IMBH orbital plane end their lives as TDEs due to Kozai-Lidov oscillations, hence illuminating the secondary SMBH/IMBH. Above a critical SMBH mass of $\approx 1.15 \times 10^8$ M$_{\odot}$, no TDE can occur for typical stars in an old stellar population since the Schwarzschild radius exceeds the tidal disruption radius. Consequently, any TDEs due to such massive SMBHs will remain dark. It follows that no TDEs should be observed in galaxies with bulges more massive than $\approx 4.15\times 10^{10}$ M$_{\odot}$, unless a lower-mass secondary SMBH or IMBH is also present. The secular mechanism for producing TDEs considered here therefore offers a useful probe of SMBH-SMBH/IMBH binarity in the most massive galaxies. We further show that the TDE rate can be $\approx 10^{-4}-10^{-3}$ yr$^{-1}$, and that most TDEs occur on $\approx 0.5$ Myr. Finally, we show that stars may be ejected with velocities up to thousands of km s$^{-1}$, which could contribute to the observed population of Galactic hypervelocity stars.
\end{abstract}


\section{Introduction}

Super Massive Black Holes (SMBHs) are located in most galactic nuclei and are fundamental building blocks in models of galaxy formation and evolution over the entire Hubble sequence \citep{kor13}. Unlike SMBHs ($M_{SMBH}\gtrsim 10^5\ \mathrm{M}_{\odot}$), the existence of intermediate-mass black holes (IMBHs), with masses $100\ \mathrm{M}_{\odot}\lesssim M_{IMBH}\lesssim 10^5\ \mathrm{M}_{\odot}$, still lacks observational evidence. IMBHs may be hosted by globular clusters \citep[e.g.][]{kruijssen13}, assuming that the empirical correlation between SMBHs and their stellar environments still holds for lower masses \citep{mer01,mer13}. According to the analysis of \citet{kiz17}, an $\approx 2,200\ \mathrm{M}_{\odot}$ IMBH could be located in the core of the Galactic globular cluster $47$ Tuc, while \citet{bau17} have claimed that $\omega$ Cen hosts a $\approx 40,000\ \mathrm{M}_{\odot}$ IMBH in its center. 

A few scenarios have been proposed for the formation of IMBHs.  One proposed mechanism involves stellar collisions in dense environments, such as the cores of massive star clusters.  This results in the runaway growth of a supra-massive star with a total mass reaching up to a few percent of the total cluster mass \citep{por00,gie15}, which is thought to subsequently collapse directly into an IMBH.  Similarly, IMBHs are thought to form from the direct collapse of massive Pop III stars \citep{mad01,wha12,woo17}. In both scenarios, IMBHs may spiral into the nucleus toward the SMBH until they form a binary system \citep*{fck17,fgk17,pet17}. This process can be mediated by inspiralling star clusters, which can deliver an IMBH close to the SMBH and cause the ejection of hypervelocity stars \citep{cap15,fra16,fck17}.  A corollary of this mechanism is that galactic nuclei can accumulate millisecond pulsars that emit in gamma-rays \citep*{fao18}.

Unfortunately, an IMBH would remain dark if not emitting due to accretion. A few bright pointlike ultra-luminous X-ray sources ($10^{39}\lesssim L_X/\mathrm{erg\ s}^{-1}\lesssim 10^{41}$) can arguably be explained by nothing other than an accreting IMBH \citep{kaa17}. Also, the distinctive gravitational wave signal emitted by a stellar mass black hole inspiralling onto an IMBH may help in spotting them in the future \citep{fgk17}. The tidal consumption of a star passing in the vicinity of an IMBH, so-called tidal disruption events (TDEs), may provide a definitive proof of the presence of IMBHs. The rate of TDEs due to SMBHs in galactic nuclei is highly uncertain and estimated to be of the order of $10^{-5}-10^{-4}$ yr$^{-1}$ per galaxy, both observationally and theoretically \citep{sto16,ale17}. The TDE rate may be enhanced due to the presence of a secondary SMBH or IMBH. \citet{lin15} studied the evolution of the distribution of stars around an SMBH binary due to the Kozai-Lidov mechanism (up to the octupole level of approximation) and found that a significant fraction of the total population of the stars surrounding the secondary SMBH can be depleted in $\approx 0.5$ Myr with a TDE rate of $\approx 10^{-2}$ yr$^{-1}$, while the rate decreases to $\approx 10^{-4}$ yr$^{-1}$ in the case of an SMBH-IMBH binary. \citep{che09,che11} studied SMBH binaries interacting with stars via three-body slingshots and showed that this process is accompanied by a burst of TDEs with rates as high as $1$ yr$^{-1}$ on a timescale $\approx 10^5$ yr in the case of a $10^7$ M$_\odot$ SMBH binary. Recently, \citet{wal18} and \citet{wang18} studied the fate of binary systems orbiting binary black holes and found an enhancement in the rate of TDE events.

In this paper, we study how stars orbiting an IMBH secondary are perturbed by an SMBH primary. We focus our attention on the TDE rate due to the tidal consumption of stars by the IMBH, driven by perturbations from the more massive SMBH. Above a critical SMBH mass of $\approx 1.15 \times 10^8$ M$_{\odot}$, no TDE event can occur for typical stars in an old stellar population (i.e., masses of $\sim$ 1 M$_{\odot}$). This is because the Schwarzschild radius exceeds the tidal disruption radius at this critical SMBH mass, such that any TDE will remain dark. From the $M_{\rm SMBH}$-$M_{\rm gal}$-relation \citep{mcconnell13}, this critical SMBH mass corresponds to a host galaxy bulge mass of $\approx 4.15\times 10^{10}$ M$_{\odot}$. This predicts that no TDEs should be observed in galaxies more massive than this critical galaxy mass, \textit{unless} a lower-mass secondary SMBH or IMBH is also present. It follows that the (secular) mechanism considered in this paper for producing TDE events by a lower mass secondary SMBH/IMBH offers a clear observational window to probe SMBH-SMBH or SMBH-IMBH binarity in the most massive galaxies. Here, we quantify the rates of such TDE events due to secular evolution of stars orbiting an SMBH-IMBH binary. 

We use high-precision direct N-body simulations, including post-Newtonian terms, to study the effects of the gravitational pull of the SMBH on the motion of stars bound to the IMBH as function of the SMBH-IMBH mass ratio and IMBH orbit. We show that the continuous perturbations exerted by the primary SMBH may lead to variations in the eccentricities and inclinations of the stars orbiting the IMBH. This results either in the ejection of the stars, or in a TDE by the IMBH, depleting its stellar surroundings.

The paper is organized as follows: in Sect. \ref{sect:met} we describe our approach to studying the secular evolution of SMBH-IMBH-star interactions; in Sect. \ref{sect:res} the results of our numerical experiments are presented and discussed. Finally, in Sect. \ref{sect:conc} we draw our conclusions.

\section{Method}
\label{sect:met}

\begin{figure} 
\centering
\includegraphics[scale=0.55]{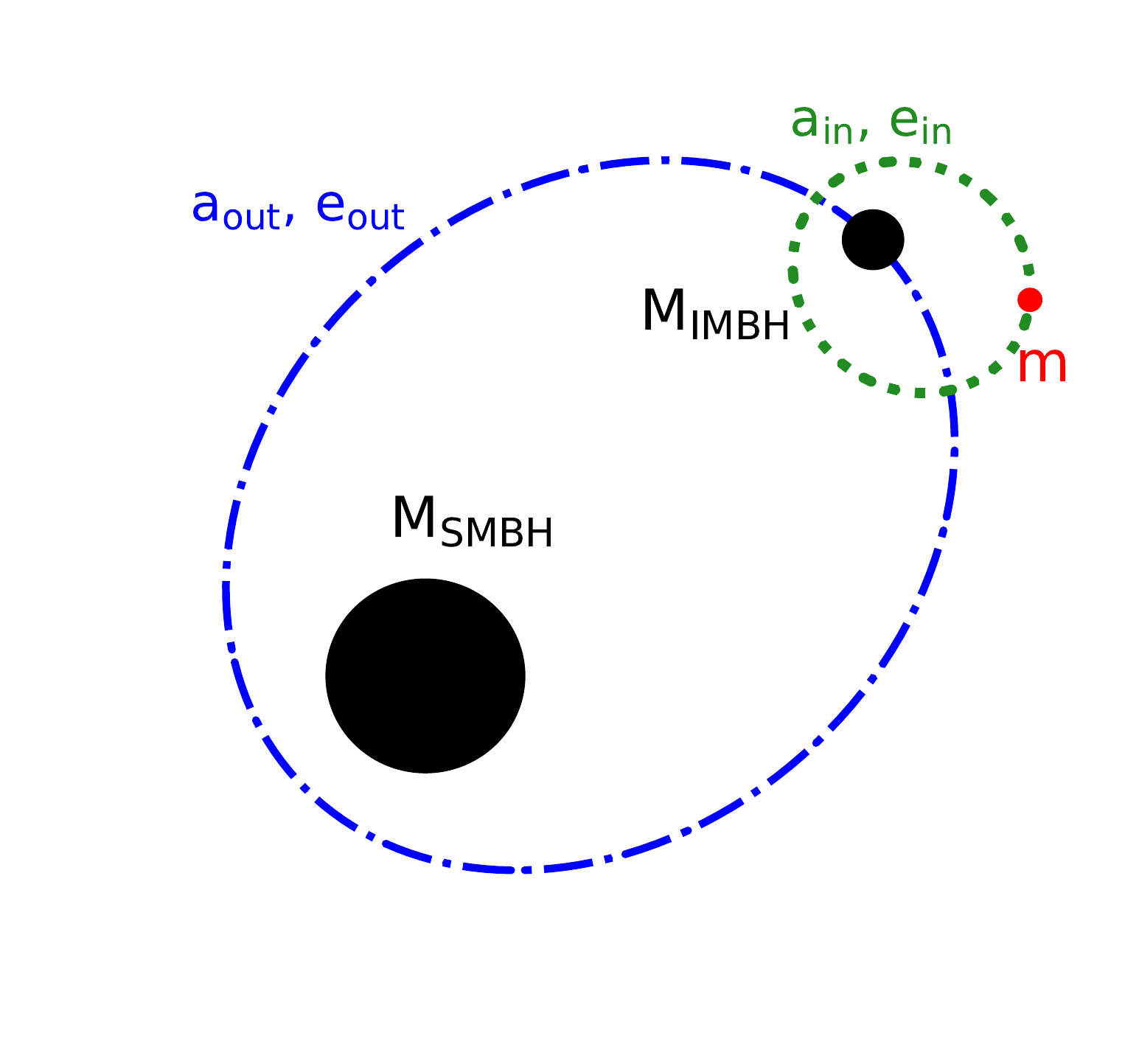}
\caption{The three-body system studied in the present work. We denote the mass of the SMBH as $M_\mathrm{SMBH}$, the mass of the secondary IMBH as $M_\mathrm{IMBH}$ and the mass of the star as $m_*$. The semimajor axis and eccentricity of the inner orbit are $a_{in}$ and $e_{in}$, respectively, while for the outer orbit $a_{in}$ and $e_{in}$.}
\label{fig:threebody}
\end{figure}

We study the fate of stars in galactic nuclei that host a massive black hole binary made up of an SMBH and an IMBH, as a function of the SMBH-IMBH mass ratio and the IMBH orbit. We consider a three-body hierarchical system consisting of an inner binary comprised of the IMBH and a star, and an outer binary comprised of the SMBH and the centre of mass of the inner IMBH-star binary. As shown in Fig. \ref{fig:threebody}, we denote the mass of the SMBH as $M_\mathrm{SMBH}$, the mass of the secondary IMBH as $M_\mathrm{IMBH}$ and the mass of the star as $m_*$. The semimajor axis and eccentricity of the inner orbit are $a_{in}$ and $e_{in}$, respectively, while for the outer orbit these are $a_{out}$ and $e_{out}$.

\begin{figure*} 
\centering
\begin{minipage}{20.5cm}
\subfloat{\includegraphics[scale=0.58]{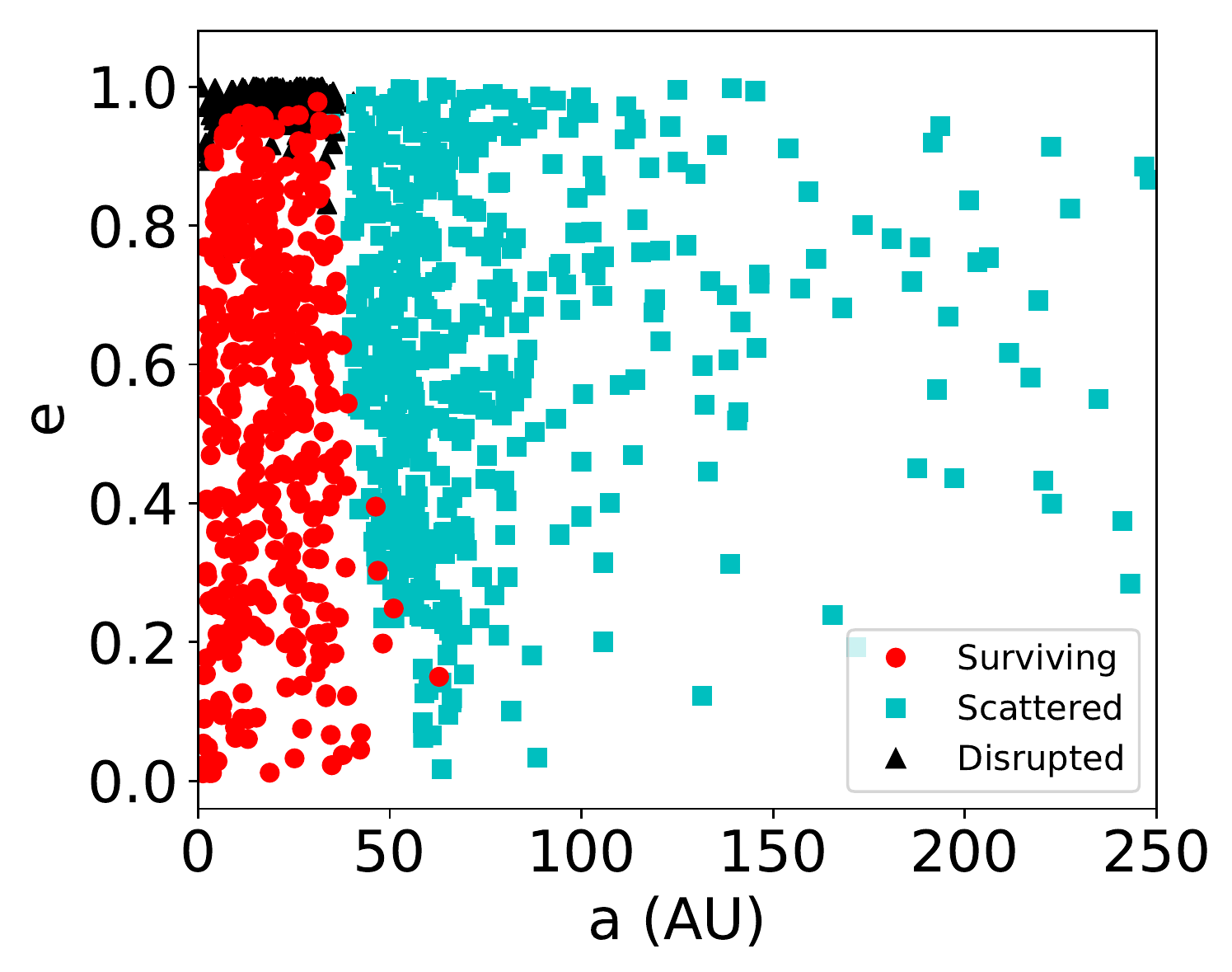}}
\subfloat{\includegraphics[scale=0.58]{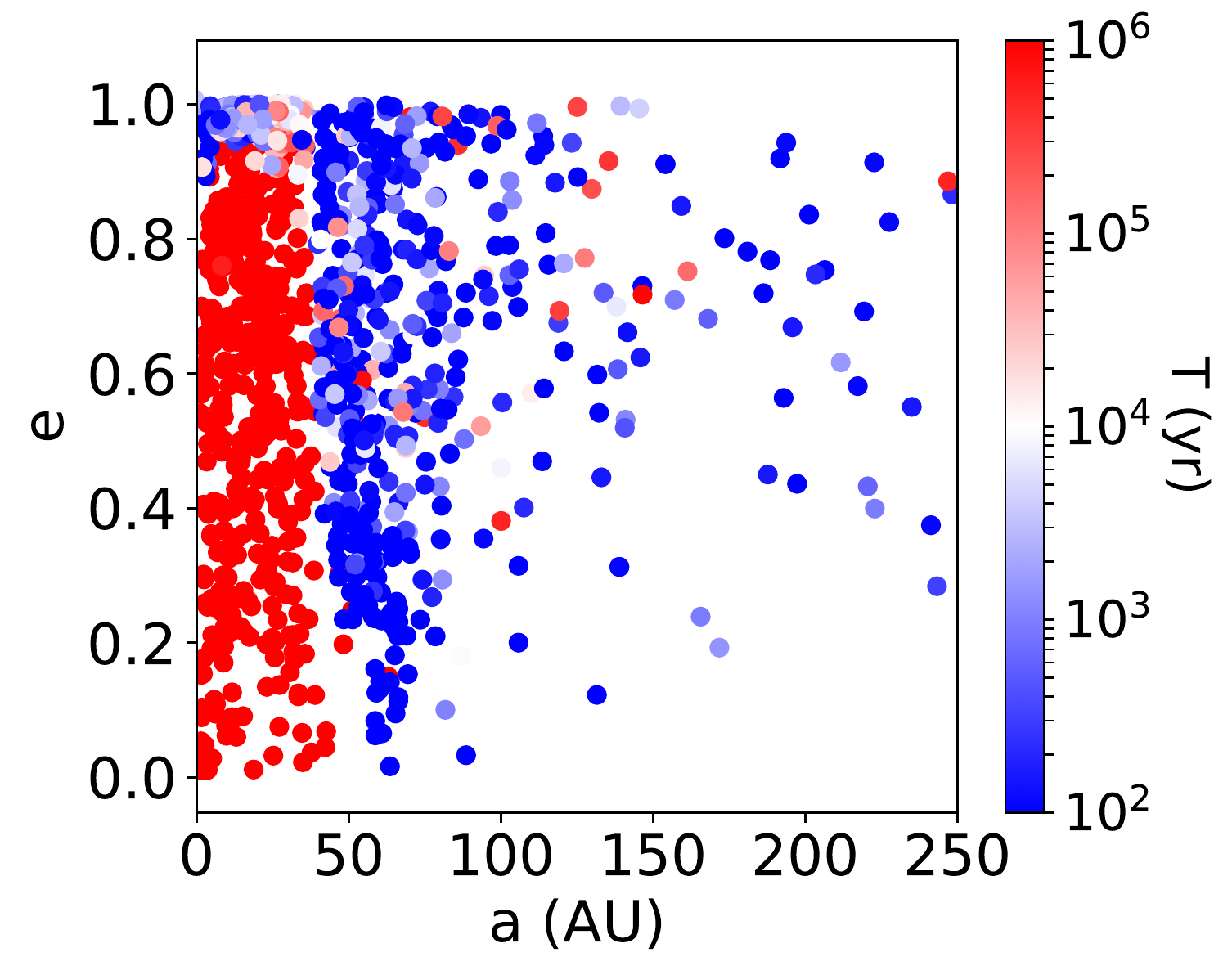}}
\end{minipage}
\caption{The outcome of the evolution of star orbits for the case of an SMBH-IMBH binary with $M_\mathrm{SMBH}=4\times 10^6$ M$_\odot$, $M_\mathrm{SMBH}=1\times 10^3$ M$_\odot$, $a_{out}=0.01$ pc and $e_{out}=0.4$. On the left, we show how different outcomes populate different regions in the semimajor axis-eccentricity-plane. On the right, we illustrate the different typical timescales for stellar disruptions and scatterings out of the Hill sphere of the IMBH.}
\label{fig:stars_final}
\end{figure*}

If the orbital plane of the star is inclined relative to the plane of the IMBH-SMBH orbit by an angle $40^\circ\lesssim i\lesssim 140^\circ$, the eccentricity and inclination of the star's orbit can experience periodic oscillations on a secular Kozai-Lidov timescale \citep{koz62,lid62}
\begin{equation}
T_{LK}=\frac{8}{15\pi}\frac{P_{out}^2}{P_{in}}\left(1-e_{out}^2\right)^{3/2}\ .
\end{equation}
In the previous equation (at the approximation of the quadrupole level), $P_{out}$ and $P_{in}$ are the periods of the outer and inner orbit, respectively. The orbital eccentricity of the star slowly increases while the inclination decreases and vice versa, conserving angular momentum \citep{nao16,grish17}. On a Kozai-Lidov timescale, the inner orbit can be excited up to a maximum eccentricity set by the initial relative inclination $i_0$ of the two orbital planes
\begin{equation}
e_{in,max}=\sqrt{1-\frac{5}{3}\cos i_0^2}\ .
\label{eqn:emax}
\end{equation}
However, Kozai-Lidov cycles can be suppressed by additional sources of apsidal precession \citep{nao16}. The most relevant process to consider is general relativistic precession that operates on a typical timescale
\begin{equation}
T_{GR}=\frac{a_{in}^{5/2}c^2(1-e_{in}^2)}{3G^{3/2}(M_{IMBH}+m_*)^{3/2}}\ .
\end{equation}
In the region of the parameter space where $T_{KL}>T_{GR}$, the Kozai-Lidov oscillations of the star orbital elements are damped by relativistic effects.

\begin{table}
\caption{Models: mass of the primary SMBH ($M_\mathrm{SMBH}$), mass of the secondary IMBH ($M_\mathrm{IMBH}$), semimajor axis of the outer orbit ($a_{out}$), eccentricity of the outer orbit ($e_{out}$).}
\centering
\begin{tabular}{cccc}
\hline
$M_\mathrm{SMBH}$	(M$_\odot$) & $M_\mathrm{IMBH}$	(M$_\odot$) & $a_{out}$	(pc) & $e_{out}$\\
\hline\hline
$4\times 10^6$ & $1$-$5$-$10\times 10^3$ & $0.01$ & $0.4$ \\
$4\times 10^6$ & $5\times 10^3$ & $0.01$-$0.05$-$0.1$ & $0.4$ \\
$4\times 10^6$ & $5\times 10^3$ & $0.01$ & $0$-$0.4$-$0.7$ \\
$1\times 10^8$ & $1$-$5$-$10\times 10^3$ & $0.01$ & $0.4$ \\
$1\times 10^8$ & $5\times 10^3$ & $0.01$-$0.05$-$0.1$ & $0.4$ \\
$1\times 10^8$ & $5\times 10^3$ & $0.01$ & $0$-$0.4$-$0.7$ \\
\hline
\end{tabular}
\label{tab:models}
\end{table}

The initial conditions for the N-body simulations have been set as follows (see also Table\,\ref{tab:models}):
\begin{itemize}
\item the mass of the SMBH is set to $M_\mathrm{SMBH}=4\times 10^6\msun$ (Milky-Way like nucleus) or $M_\mathrm{SMBH}=10^8\msun$;
\item the mass of the IMBH is set to $M_\mathrm{IMBH}=10^3\msun$-$5\times 10^3\msun$-$10^4\msun$;
\item the mass of the star orbiting the IMBH is fixed to $m_*=1\msun$;
\item the semimajor axis of the SMBH-IMBH orbit is $\aout=0.01\au$-$0.05\au$-$0.1\au$;
\item the eccentricity of the SMBH-IMBH orbit is $e_\mathrm{out}=0$-$0.4$-$0.7$;
\item the semimajor axis of the star orbiting the IMBH is sampled uniformly within the Hill sphere of the IMBH at its orbital pericentre with respect to the SMBH
\begin{equation}
R_{H}=a_{out}(1-e_{out})\left(\frac{M_{IMBH}}{M_{SMBH}}\right)^{1/3}
\label{eqn:hills}
\end{equation}
\item the eccentricity of the star is sampled from a uniform distribution;
\item the mutual inclination $i$ between the inner and outer orbital planes is drawn from an isotropic distribution;
\item the initial phase $\Psi$ of the IMBH-star centre of mass orbit around the SMBH and the initial phase $\Phi$ of the star orbit around the IMBH are chosen randomly.
\end{itemize}

Given the above set of initial parameters, we integrate the system of differential equations of motion of the 3-bodies
\begin{equation}
{\ddot{\textbf{r}}}_i=-G\sum\limits_{j\ne i}\frac{m_j(\textbf{r}_i-\textbf{r}_j)}{\left|\textbf{r}_i-\textbf{r}_j\right|^3}\ ,
\end{equation}
with $i=1$,$2$,$3$.  The integrations are performed using the \textsc{ARCHAIN} code \citep{mik06,mik08}, a fully regularized code able to model the evolution of binaries of arbitrary mass ratios and eccentricities with extreme accuracy, even over long periods of time. \texttt{ARCHAIN} includes PN corrections up to order PN2.5. 

Previous works adopted the secular approximation to study the fate of stars orbiting binary massive black holes. Although faster and less demanding from a computational point of view, the secular approximation fails to describe properly the equations of motion when the perturbation exerted by the outer SMBH is strong or the eccentricity is close to unity \citep{anm14,lin15,nao16}. For this reason, we use high-precision direct N-body simulations to study accurately the long term evolution of the three-body system.

\begin{figure} 
\centering
\includegraphics[scale=0.58]{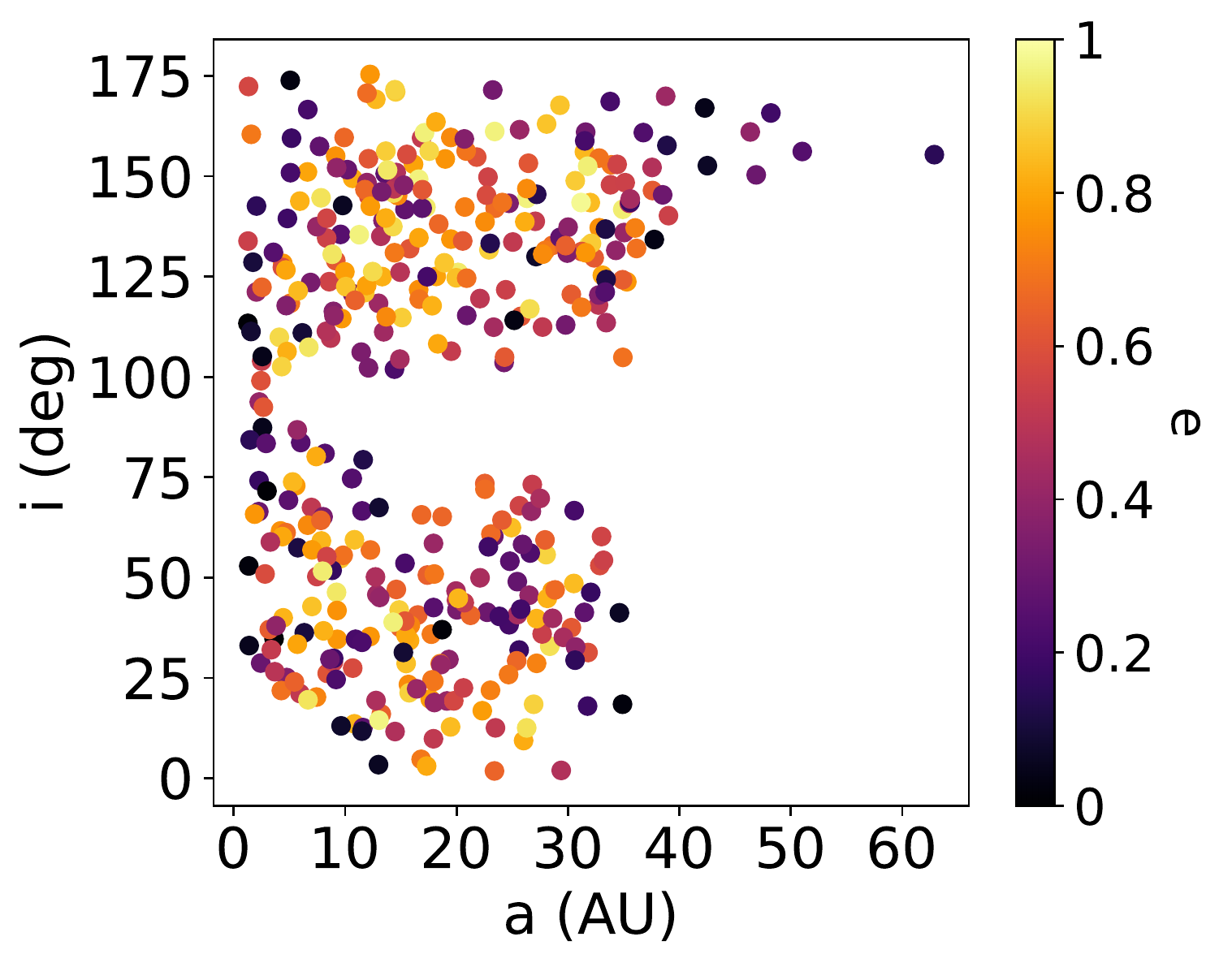}
\caption{The final distribution of surviving stars after $1$ Myr in the case of an SMBH-IMBH binary with $M_\mathrm{SMBH}=4\times 10^6$ M$_\odot$, $M_\mathrm{SMBH}=1\times 10^3$ M$_\odot$, $a_{out}=0.01$ pc and $e_{out}=0.4$ in the semimajor axis-inclination-space. Out to a few AU's from the IMBH, surviving stars are distributed in the shape of a torus. Stars in the innermost region have less extreme eccentricities than stars in the outer part and are distributed in a more spherical configuration.}
\label{fig:sem_incl_ecc}
\end{figure}

\section{Results}
\label{sect:res}

\begin{figure*} 
\centering
\begin{minipage}{20.5cm}
\subfloat{\includegraphics[scale=0.58]{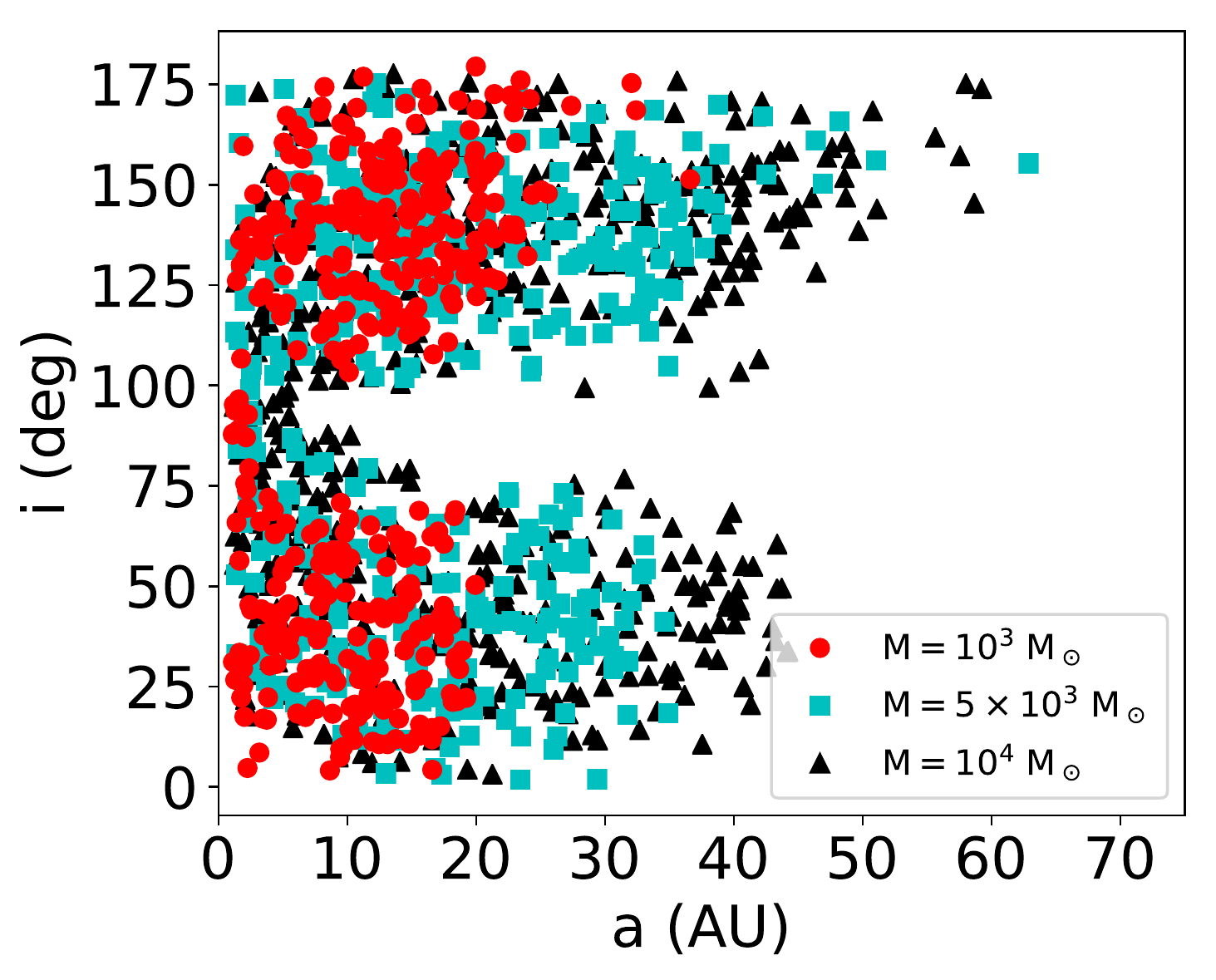}}
\subfloat{\includegraphics[scale=0.58]{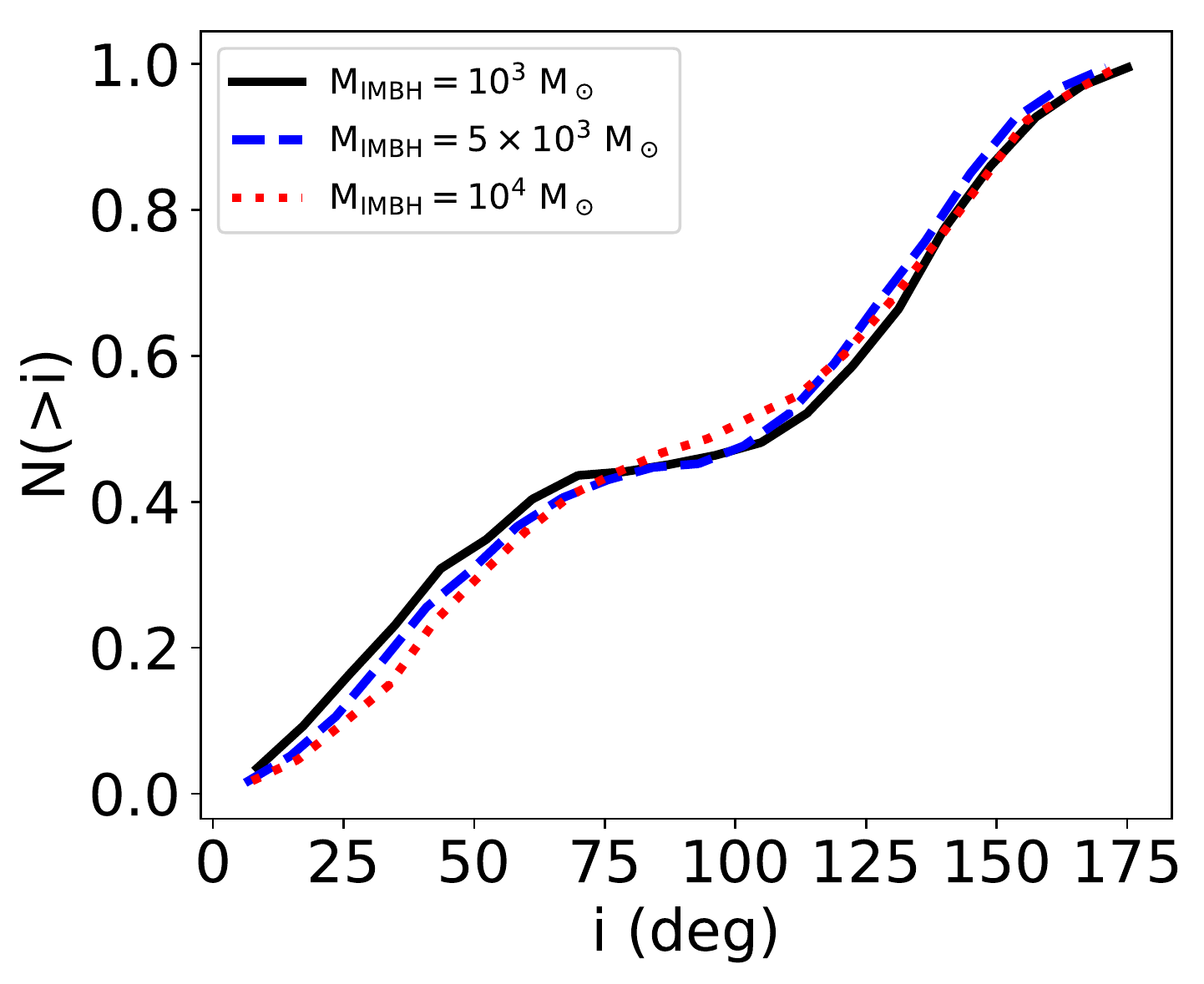}}
\end{minipage}
\caption{Left panel: the final distribution of surviving stars after $1$ Myr in the case of an SMBH-IMBH binary with $M_\mathrm{SMBH}=4\times 10^6$ M$_\odot$, $a_{out}=0.01$ pc and $e_{out}=0.4$ and different IMBH masses. Out to a few AU's from the IMBH, surviving stars are distributed in the shape of a torus. Right panel: the final cumulative distribution of the inclinations of surviving stars after $1$ Myr.}
\label{fig:stars_ecc_incl}
\end{figure*}

\begin{figure*} 
\centering
\begin{minipage}{20.5cm}
\subfloat{\includegraphics[scale=0.55]{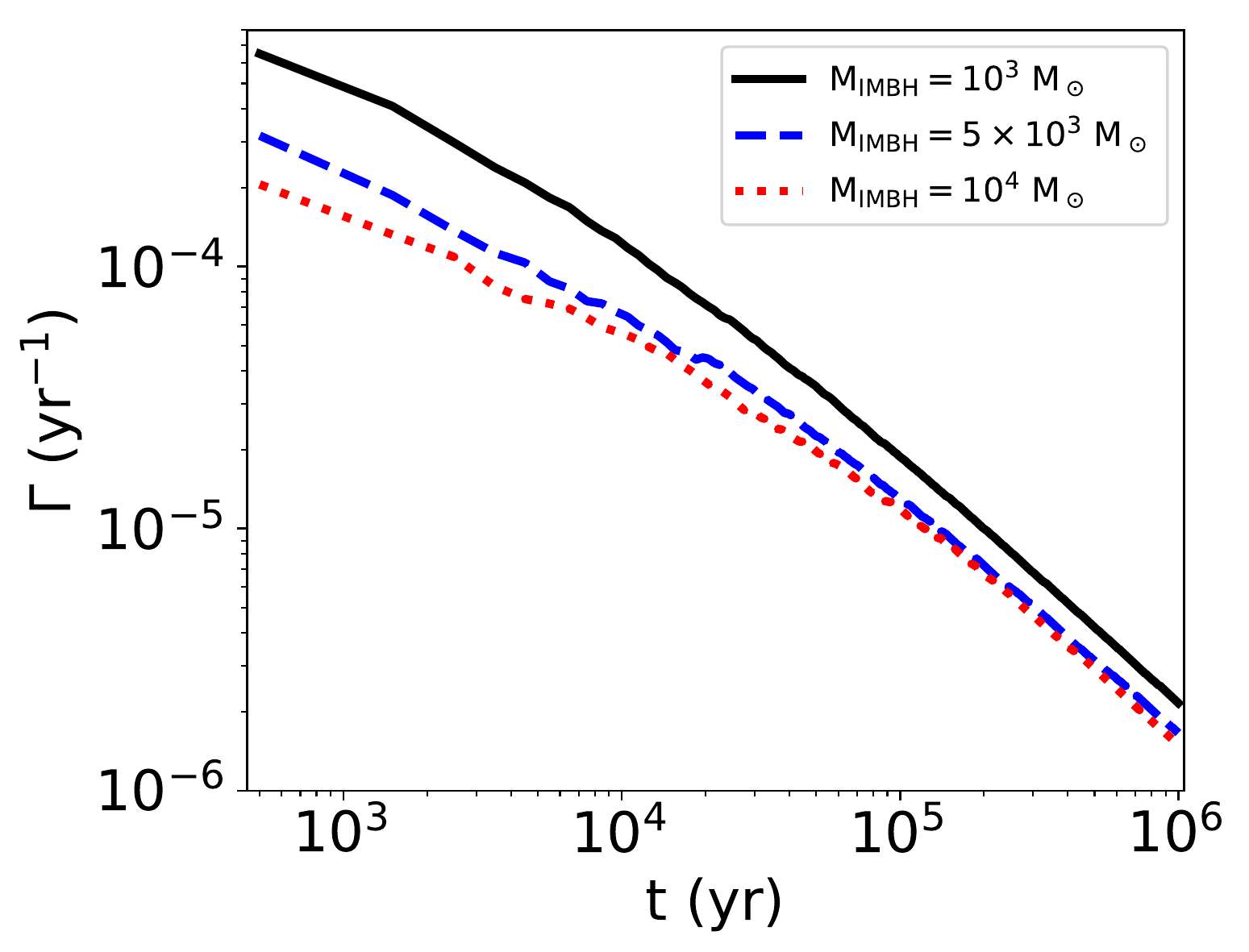}}
\subfloat{\includegraphics[scale=0.55]{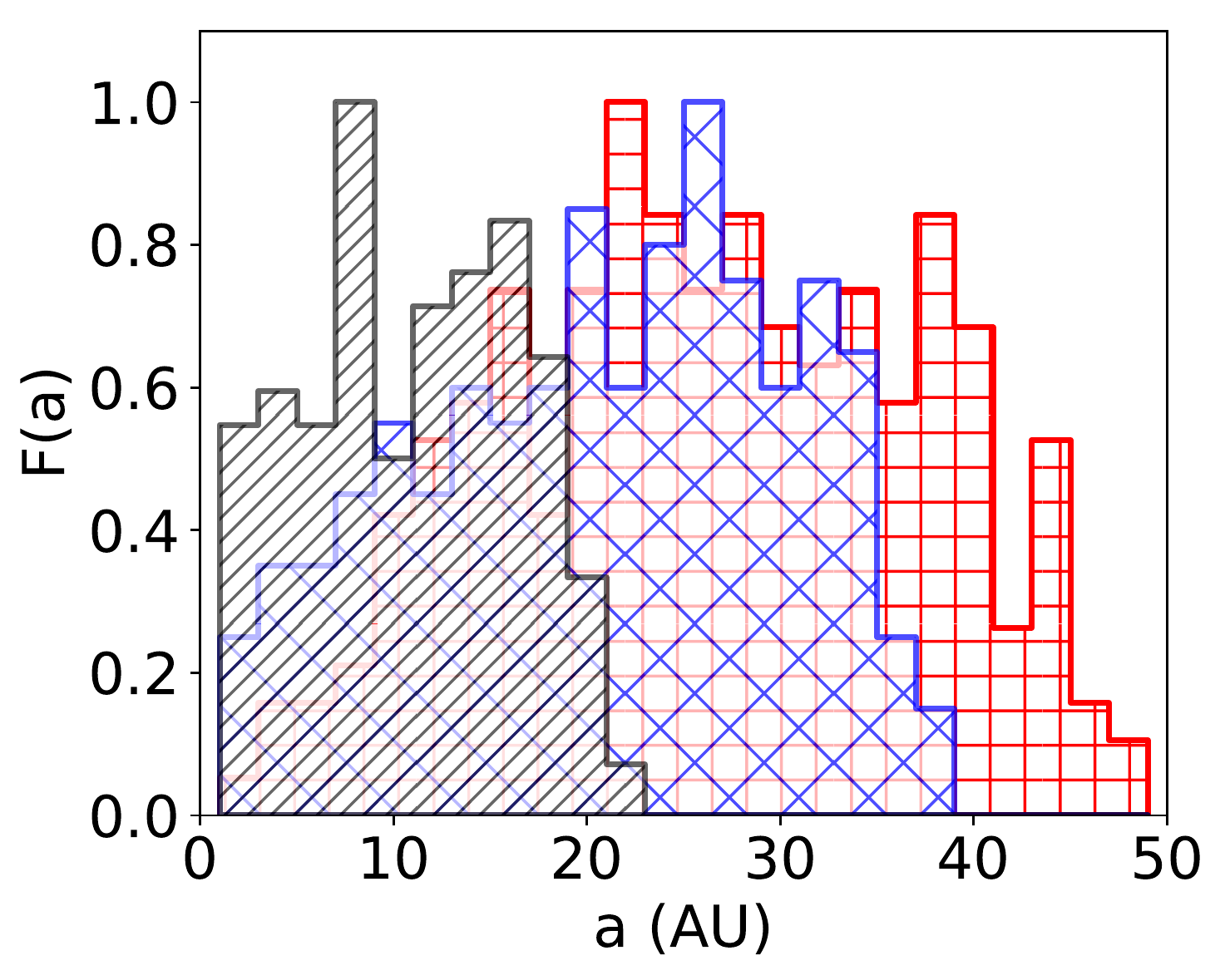}}
\end{minipage}
\begin{minipage}{20.5cm}
\subfloat{\includegraphics[scale=0.55]{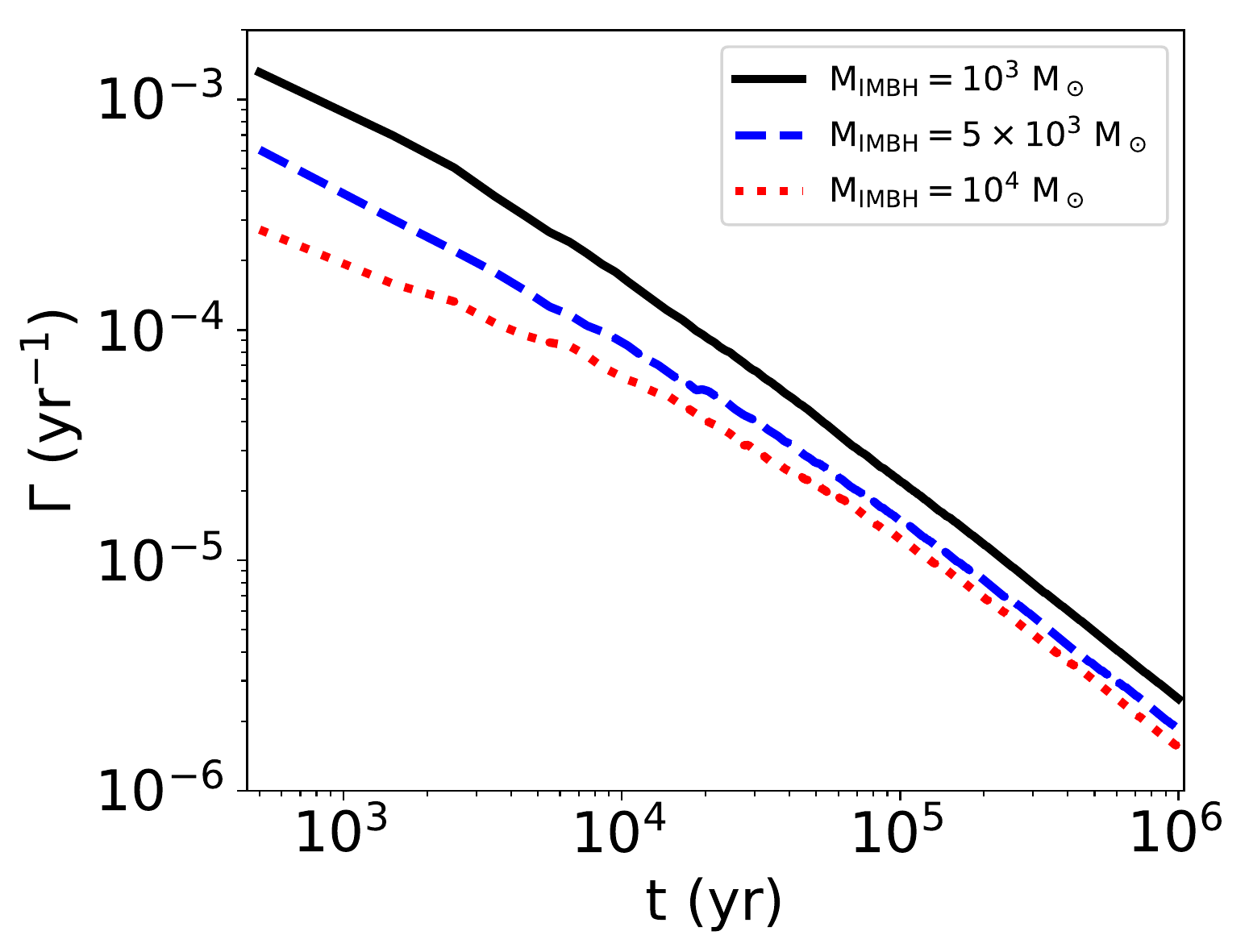}}
\subfloat{\includegraphics[scale=0.55]{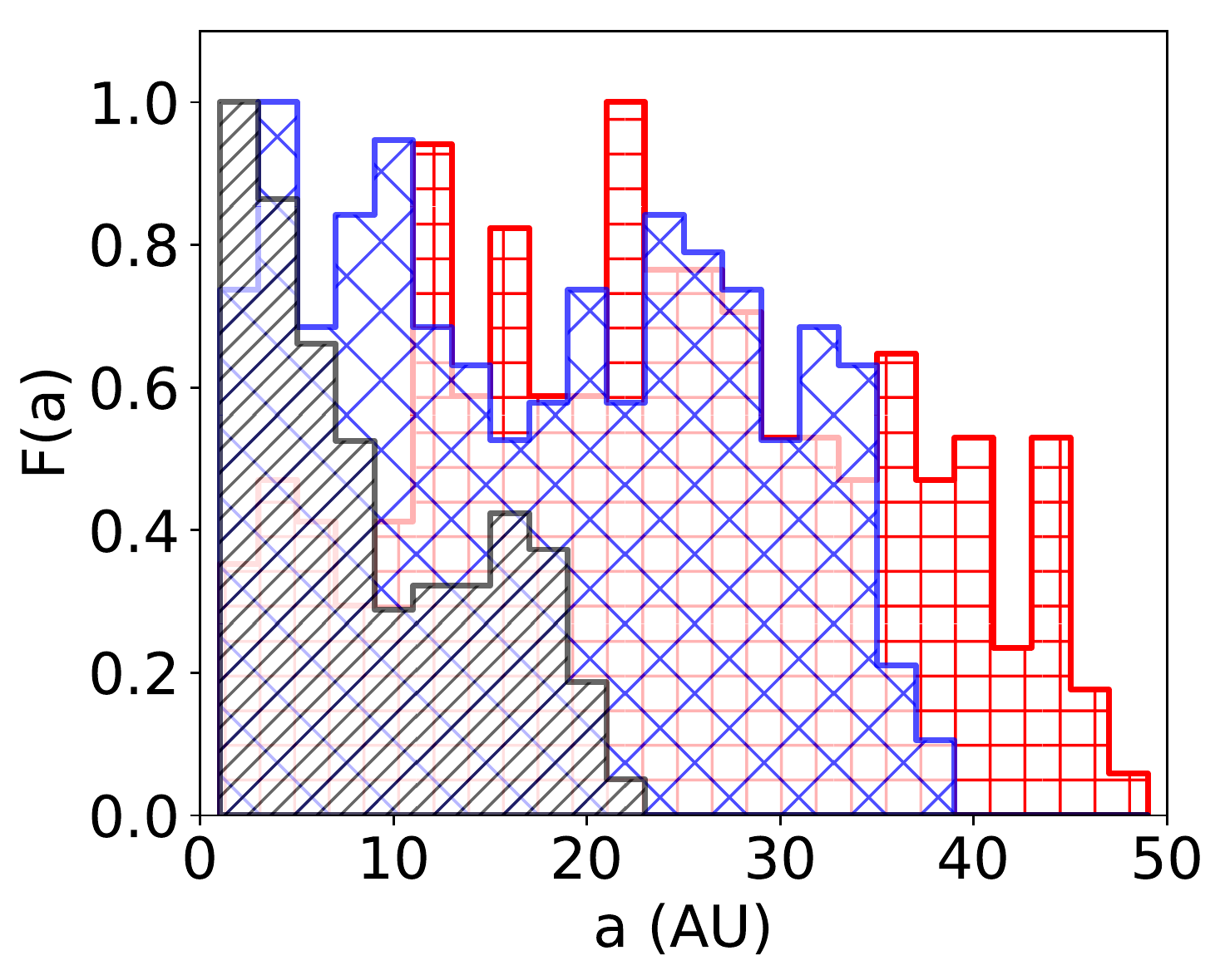}}
\end{minipage}
\caption{Left: TDE rate as function of time and semimajor axis of tidally disrupted stars in the case of an SMBH-IMBH binary with $M_\mathrm{SMBH}=4\times 10^6$ M$_\odot$ (top) and $M_\mathrm{SMBH}=4\times 10^6$ M$_\odot$ (bottom), $a_{out}=0.01$ pc, $e_{out}=0.4$ and different IMBH masses. Right: distribution of semimajor axis of stars that end their lives as TDEs. Linestyle as in the right panel.}
\label{fig:rate_tde}
\end{figure*}

\begin{figure*} 
\centering
\begin{minipage}{20.5cm}
\subfloat{\includegraphics[scale=0.425]{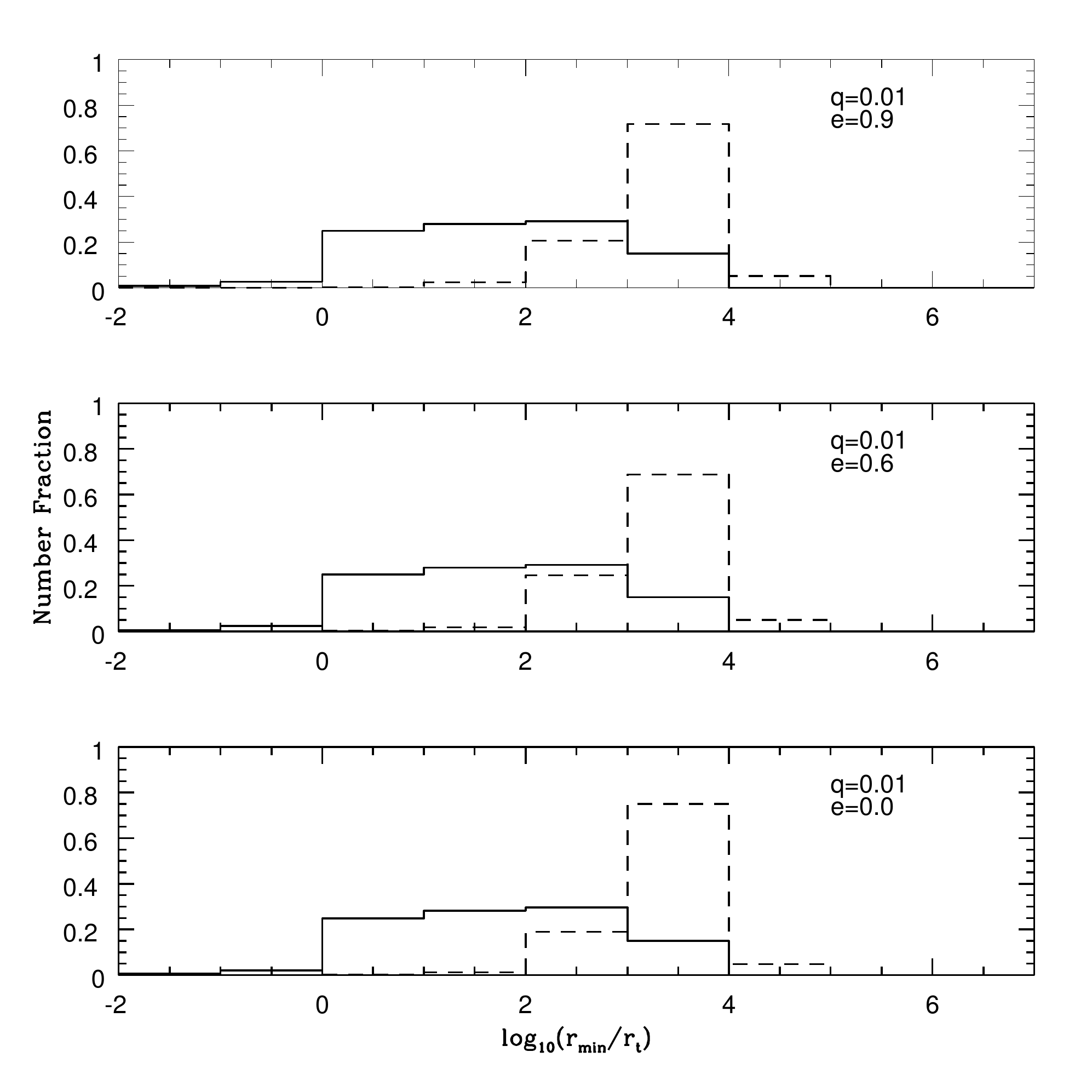}}
\subfloat{\includegraphics[scale=0.425]{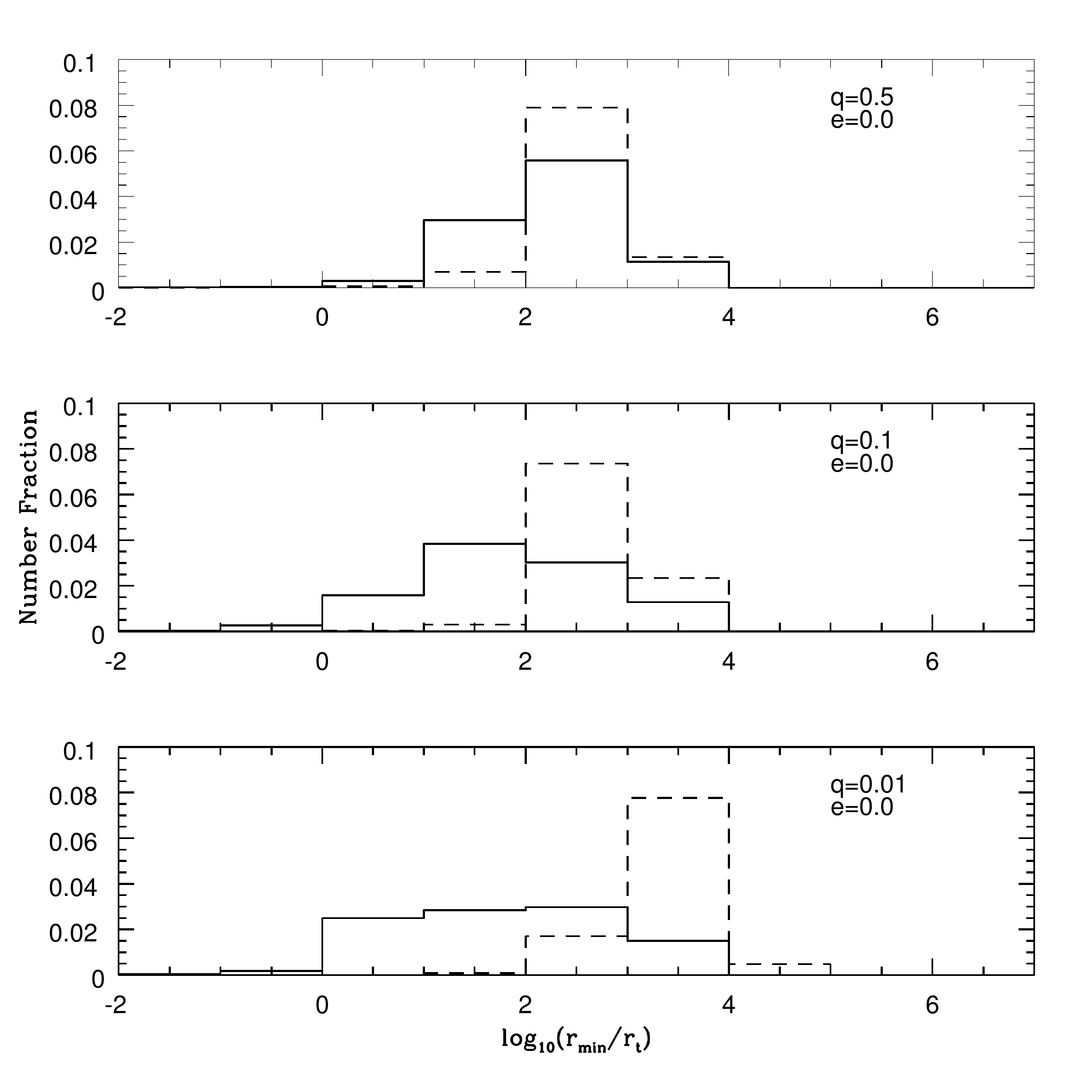}}
\end{minipage}
\caption{Numerical scattering simulations for the tidal disruption of stars via dynamical scattering interactions, performed using the \texttt{FEWBODY} code. The initial conditions for these simulations are described in the text. The solid lines show the distance of closest approach of the star relative to the primary SMBH, in units of its tidal disruption radius. The dashed lines show the same thing but for the secondary less massive SMBH. Left: the top, middle and bottom panels show, respectively, the results assuming binary mass ratios $q = 0.5, 0.1, 0.01$, and an orbital eccentricity of zero. Right: same as left panel, but varying the binary orbital separation at constant mass ratio; the top, middle and bottom panels show, respectively, the results assuming orbital eccentricities of $e = 0.9, 0.6, 0.0$.}
\label{fig:tdeq12}
\end{figure*}

For each set of parameters, we run $1000$ simulations up to a maximum time $T=1$ Myr. In our N-body integrations the star orbiting the IMBH has three possible fates: (i) the star can remain bound to the IMBH, but on an orbit perturbed with respect to the original one, (ii) the star can be captured by the SMBH or entirely ejected from the galactic nucleus, with velocities up to a few thousand km s$^{-1}$, or (iii) the star is tidally disrupted by the IMBH, producing a TDE event. The tidal radius for the star orbiting the IMBH is
\begin{equation}
R_t\approx r_* \left(\frac{M_{IMBH}}{M_*}\right)^{1/3}\ .
\end{equation}
As assumed in \citet{lin15}, we identify TDE events with the condition
\begin{equation}
a_{in}(1-e_{in})<3 R_t\ ,
\label{eqn:disr}
\end{equation}
since the stars may still be disrupted due to the accumulated heating by strong tides exerted outside the tidal radius \citep{lil13}. The distinction among these three possible outcomes is made by computing the mechanical energy of the star with respect to the IMBH at every integration time-step. If its energy with respect to the IMBH remains negative, the star remains bound to the IMBH (case (i)), while if this energy becomes positive, the star becomes unbound from the IMBH (case (ii)). Finally, if Eq. \ref{eqn:disr} is satisfied, the star is considered tidally accreted onto the IMBH (case (iii)).

The left panel of Fig. \ref{fig:stars_final} shows the final distribution of stars in the semimajor axis-eccentricity-plane in the case of an SMBH-IMBH binary with $M_\mathrm{SMBH}=4\times 10^6$ M$_\odot$, $M_\mathrm{SMBH}=1\times 10^3$ M$_\odot$, $a_{out}=0.01$ pc and $e_{out}=0.4$. The semimajor axis of the disrupted and scattered stars refers to the relative semimajor axis of the stars before disruption or scattering, respectively. Scattered, surviving and disrupted stars populate different regions of this parameter space. 

Above $\approx 50$ AU, there are no stable orbits around the IMBH and stars get stripped because of the continuous gravitational perturbations by the SMBH. Stars that are disrupted and undergo a TDE event onto the IMBH have smaller semimajor axis ($\lesssim 50$ AU) and high eccentricities. Stars that end their lives as TDEs orbit in a plane highly inclined with respect to the SMBH-IMBH orbital plane. As a consequence, the Kozai-Lidov mechanism starts affecting efficiently the dynamics of the system and modulates the oscillations in eccentricity and inclination, whenever not suppressed by relativistic precession, exciting the star's eccentricity to almost unity (see Eq. \ref{eqn:emax}). Stars that remain bound to the IMBH have semimajor axes $\lesssim 50$ AU, moderate eccentricities and orbital inclinations out of the Kozai-Lidov window (see Fig. \ref{fig:sem_incl_ecc}). The right panel of Fig. \ref{fig:stars_final} illustrates the typical timescales for different outcomes. While stars are typically scattered out of the IMBH's Hill sphere on short timescales ($\lesssim 10^4$ yr), stars are scattered onto TDE orbits on longer timescales $\approx 10^5$ yr.

Figure \ref{fig:stars_ecc_incl} shows the distribution of surviving stars in the semimajor axis-inclination-plane (left panel) and the cumulative distribution of the inclinations of the stellar orbits with respect to the SMBH-IMBH orbital plane, in the case of an SMBH-IMBH binary with $M_\mathrm{SMBH}=4\times 10^6$ M$_\odot$, $a_{out}=0.01$ pc and $e_{out}=0.4$ and different IMBH masses. Independent of the IMBH mass, surviving stars are distributed in a torus-like shape, apart from a spherical innermost region. The more massive the IMBH, the more extended the distribution of surviving stars. Stars orbiting in planes highly inclined with respect to the SMBH-IMBH binary cannot survive due to Kozai-Lidov cycles and are subjected to high excitations of the orbital eccentricity, which either scatters them out of the IMBH potential well or generates a TDE \citep{lin15}. The cumulative distribution of the inclinations of surviving stars illustrates that roughly half of the stars have inclinations $\lesssim 60^\circ$ and the other half have $\gtrsim 120^\circ$, approximatively near the edge or out of the Kozai-Lidov window, making them safe against high excitations in the eccentricity.

\subsection{TDE rates}

\begin{table*}
\caption{TDE rates: mass of the primary SMBH ($M_\mathrm{SMBH}$), mass of the secondary IMBH ($M_\mathrm{IMBH}$), semimajor axis of the outer orbit ($a_{out}$), eccentricity of the outer orbit ($e_{out}$), branching ratio of disrupted stars (BR1), branching ratio of scattered stars (BR2), branching ratio of surviving stars (BR3), TDE rate within $1$ Myr ($\Gamma$).} 
\centering
\begin{tabular}{cccccccc}
\hline
$M_\mathrm{SMBH}$	(M$_\odot$) & $M_\mathrm{IMBH}$	(M$_\odot$) & $a_{out}$	(pc) & $e_{out}$ & BR1 & BR2 & BR3 & $\Gamma$ (yr$^{-1}$) \\
\hline\hline
$4\times 10^6$ & $1\times 10^3$  & $0.01$ & $0.4$ & $0.239$ & $0.419$ & $0.342$ & $2.4\times 10^{-6}$ \\
$4\times 10^6$ & $5\times 10^3$  & $0.01$ & $0.4$ & $0.187$ & $0.454$ & $0.358$ & $1.9\times 10^{-6}$ \\
$4\times 10^6$ & $1\times 10^4$  & $0.01$ & $0.4$ & $0.178$ & $0.441$ & $0.381$ & $1.8\times 10^{-6}$ \\
$4\times 10^6$ & $5\times 10^3$  & $0.01$ & $0.4$ & $0.187$ & $0.454$ & $0.358$ & $1.9\times 10^{-6}$ \\
$4\times 10^6$ & $5\times 10^3$  & $0.05$ & $0.4$ & $0.092$ & $0.447$ & $0.461$ & $0.9\times 10^{-7}$ \\
$4\times 10^6$ & $5\times 10^3$  & $0.1$  & $0.4$ & $0.073$ & $0.393$ & $0.534$ & $0.7\times 10^{-6}$ \\
$4\times 10^6$ & $5\times 10^3$  & $0.01$ & $0$   & $0.203$ & $0.307$ & $0.490$ & $2.0\times 10^{-6}$ \\
$4\times 10^6$ & $5\times 10^3$  & $0.01$ & $0.4$ & $0.187$ & $0.454$ & $0.358$ & $1.9\times 10^{-6}$ \\
$4\times 10^6$ & $5\times 10^3$  & $0.01$ & $0.7$ & $0.216$ & $0.458$ & $0.326$ & $2.2\times 10^{-6}$ \\
$1\times 10^8$ & $1\times 10^3$  & $0.01$ & $0.4$ & $0.287$ & $0.423$ & $0.290$ & $2.9\times 10^{-6}$ \\
$1\times 10^8$ & $5\times 10^3$  & $0.01$ & $0.4$ & $0.214$ & $0.460$ & $0.326$ & $2.1\times 10^{-6}$ \\
$1\times 10^8$ & $1\times 10^4$  & $0.01$ & $0.4$ & $0.177$ & $0.453$ & $0.370$ & $1.8\times 10^{-6}$ \\
$1\times 10^8$ & $5\times 10^3$  & $0.01$ & $0.4$ & $0.214$ & $0.460$ & $0.326$ & $2.1\times 10^{-6}$ \\
$1\times 10^8$ & $5\times 10^3$  & $0.05$ & $0.4$ & $0.111$ & $0.497$ & $0.392$ & $1.1\times 10^{-6}$ \\
$1\times 10^8$ & $5\times 10^3$  & $0.1$  & $0.4$ & $0.103$ & $0.482$ & $0.415$ & $1.0\times 10^{-6}$ \\
$1\times 10^8$ & $5\times 10^3$  & $0.01$ & $0$   & $0.339$ & $0.179$ & $0.482$ & $3.4\times 10^{-6}$ \\
$1\times 10^8$ & $5\times 10^3$  & $0.01$ & $0.4$ & $0.214$ & $0.460$ & $0.326$ & $2.1\times 10^{-6}$ \\
$1\times 10^8$ & $5\times 10^3$  & $0.01$ & $0.7$ & $0.302$ & $0.411$ & $0.287$ & $3.0\times 10^{-6}$ \\
\hline
\end{tabular}
\label{tab:br}
\end{table*}

As discussed in the previous section, stars whose orbital inclination are within the Kozai-Lidov window ($40^\circ\lesssim i \lesssim 140^\circ$) suffer from high excitations both in eccentricity and inclination and may end their lifetime as TDEs. Table \ref{tab:br} reports the final branching ratios (BRs) for the three different outcomes and the inferred rate $\Gamma$ of TDEs within $1$ Myr for all the parameters considered in the present work. The probability of the different outcomes depends on the mass ratio of the SMBH-IMBH binary and on the IMBH orbital semimajor axis and eccentricity. We find that the smaller the mass ratio between the SMBH and IMBH, the larger the probability for stars to be tidally disrupted and the smaller the probability for stars to survive on a bound orbit around the IMBH. Larger semimajor axes enhance the probability of stars remaining bound to the IMBH (on perturbed orbits with respect to the initial one) and the average probabilities to undergo a TDE event or to be scattered off the IMBH's Hill sphere become smaller. Finally, larger IMBH orbital eccentricities imply a larger probability for the stars to be scattered and a smaller probability of remain bound to the IMBH, while the TDE channel remains nearly insensitive to the eccentricity. 

We have also computed the TDE rate for all the models considered in this work, by considering that an IMBH may still have $N_* \approx 10$ stars orbiting it upon arriving in a galactic nucleus. $N_*$ can be calculated by requiring that the typical timescale for star collisions ($T_{coll}$) is larger than the timescale of interest for TDEs. Most of the TDEs occur on a timescale of $\approx 0.5$ Myr, which, as noted by \citet{lin15}, corresponds roughly to the Kozai-Lidov time-scale at the octupole level of approximation. For instance, assuming $M_{IMBH}=10^3 \msun$ and as maximum extent of the IMBH sphere of influence $\approx 30$ AU
\begin{equation}
T_{coll}\approx \frac{6 \times 10^6}{N_*}\ \mathrm{yr} \gtrsim 0.5\ \mathrm{Myr}\ , 
\end{equation}
which implies $N_*\lesssim 12$. As discussed, the smaller the mass ratio the larger the TDE outcome probability. Figure \ref{fig:rate_tde} (left panel) shows the TDE rate as a function of time and the semimajor axis of tidally disrupted stars in the case of an SMBH-IMBH binary with $M_\mathrm{SMBH}=4\times 10^6$ M$_\odot$ (top) and $M_\mathrm{SMBH}=10^8$ M$_\odot$ (bottom), $a_{out}=0.01$ pc, $e_{out}=0.4$ and different IMBH masses. Within $1$ Myr, the TDE rate can be as high as $\approx 10^{-4}-10^{-3}$ yr$^{-1}$. Figure \ref{fig:rate_tde} (right panel) shows that the larger the mass ratio the larger the average semimajor axis of stars that can undergo TDEs.

We note that our study takes into account what is the fate of the stars still bound to the IMBH when the IMBH has a long-term stationary orbit around the SMBH and do not take into account the process that delivers the IMBH to the center of the galactic nucleus. Probably, two main physical scenarios can bring the IMBH to the inner nucleus, i.e. dynamical friction acting on an isolated IMBH and dynamical friction acting on the host star cluster to an IMBH \citep{fgk17,agu18}. In both cases, the typical timescale for the IMBH to reach a stable orbit about the SMBH ranges from a few Myrs to tens of Myrs, but these estimates are sensitive to the initial conditions (e.g., the mass of the IMBH, the mass of the host star cluster, the pericenter of the IMBH's orbit, etc.). We do not take into account the previous dynamical history that delivers the IMBH to this stable orbit, and instead focus on the fate of its bound stars once this steady-state is reached. However, if we assume a typical timescale of $\approx 10$ Myr to deliver an IMBH to the inner galactic nucleus, our rate would decrease by a factor of $\approx 10$. Finally, we note that the TDE rate may be enhanced to very high rates during the complex process of delivering the IMBH to this position, due to chaotic N-body interactions between the IMBH and its immediate stellar environment.

We emphasize that, since this mechanism for producing TDEs by an IMBH secondary becomes more efficient at lower mass ratios, it offers an ideal observational test to probe SMBH-SMBH or SMBH-IMBH binarity in the most massive galaxies in this regime of mass ratios.  This is because, as discussed in the Introduction, above a critical SMBH mass of $\approx 1.15 \times 10^8$ M$_{\odot}$, no TDE event can occur (assuming masses and radii of $1$ M$_{\odot}$ and $1$ R$_{\odot}$ for the stars). This is because the Schwarzschild radius exceeds the tidal disruption radius at this critical SMBH mass, such that any TDE will remain dark. From the $M_{\rm SMBH}$-$M_{\rm gal}$-relation \citep{mcconnell13}, this critical SMBH mass corresponds to a host galaxy bulge mass of $\sim$ 4.15 $\times$ 10$^{10}$ M$_{\odot}$. This predicts that no TDEs should be observed in galaxies more massive than this critical galaxy mass, \textit{unless} a lower-mass secondary SMBH or IMBH is also present.\footnote{We caution that this ignores the spin of the SMBH, which can increase this maximum mass limit \citep{beloborodov92,kesden12}.} We have shown in this paper that the probability of a TDE event initiated by the much lower-mass secondary is high in the secular limit.

To help illustrate that the secular evolution considered in this paper is a more promising mechanism for detecting SMBH secondaries in the most massive galaxies relative to the dynamical scattering of stars onto loss-cone orbits, we have performed numerical scattering simulations using the \texttt{FEWBODY} code \citep{fregeau04}. For these simulations, we fix the mass of the primary to be $M_{\rm 1} =$ 10$^6$ M$_{\odot}$, but vary the mass of the secondary.  We sample mass ratios $q = 0.01, 0.1, 0.5$, where $q = M_{\rm 2}/M_{\rm 1}$.  We assume circular orbits for the SMBH-SMBH binaries for $q = 0.1, 0.5$ (see Figure~\ref{fig:tdeq12} left panel), but consider different eccentricities for the mass ratio $q = 0.01$, namely $e = 0.0, 0.6, 0.9$ (see Figure~\ref{fig:tdeq12} right panel).  We sample the relative velocity at infinity for the incoming single star in the range $3\times 10^{-3} (M_{\rm m2}/M)^{1/2} < v/v_{\rm c} < 30(M_{\rm 2}/M)$ with $80$ logarithmically equally spaced grid points, following \citet{sesana06}. Here, $M = M_{\rm 1} + M_{\rm 2}$ is the total mass of the SMBH-SMBH binary, $v_{\rm c} = (GM/a)^{1/2}$ is the critical velocity and $a$ is the initial binary semi-major axis. We fix $a = 1$ AU, but note that varying the incoming velocity at infinity is equivalent to varying the binary semi-major axis $a$ at fixed $v$. For each incoming velocity, we sample the impact parameter $b$ randomly according to an equal probability distribution in $b^2$, within the interval corresponding to a range in scaled pericentre distance $r_{\rm p}/a$ of [0,5]. Finally, the two velocity angles are randomly generated to reproduce an uniform density distribution over a spherical surface centered on the origin of the coordinate system, while the binary phase orientation angle is chosen from a uniform distribution in the range [0,2$\pi$]. For each binary mass ratio and eccentricity, we perform a total of $10^6$ numerical scattering experiments. We have also verified consistency with the results in \citet{sesana06} for the range of binary mass ratios and eccentricities considered here. Lower mass ratios and higher eccentricities will be the subject of a following paper and should use the \texttt{ARchain} integration scheme to ensure accurate integrations.

\begin{figure*} 
\centering
\begin{minipage}{20.5cm}
\subfloat{\includegraphics[scale=0.55]{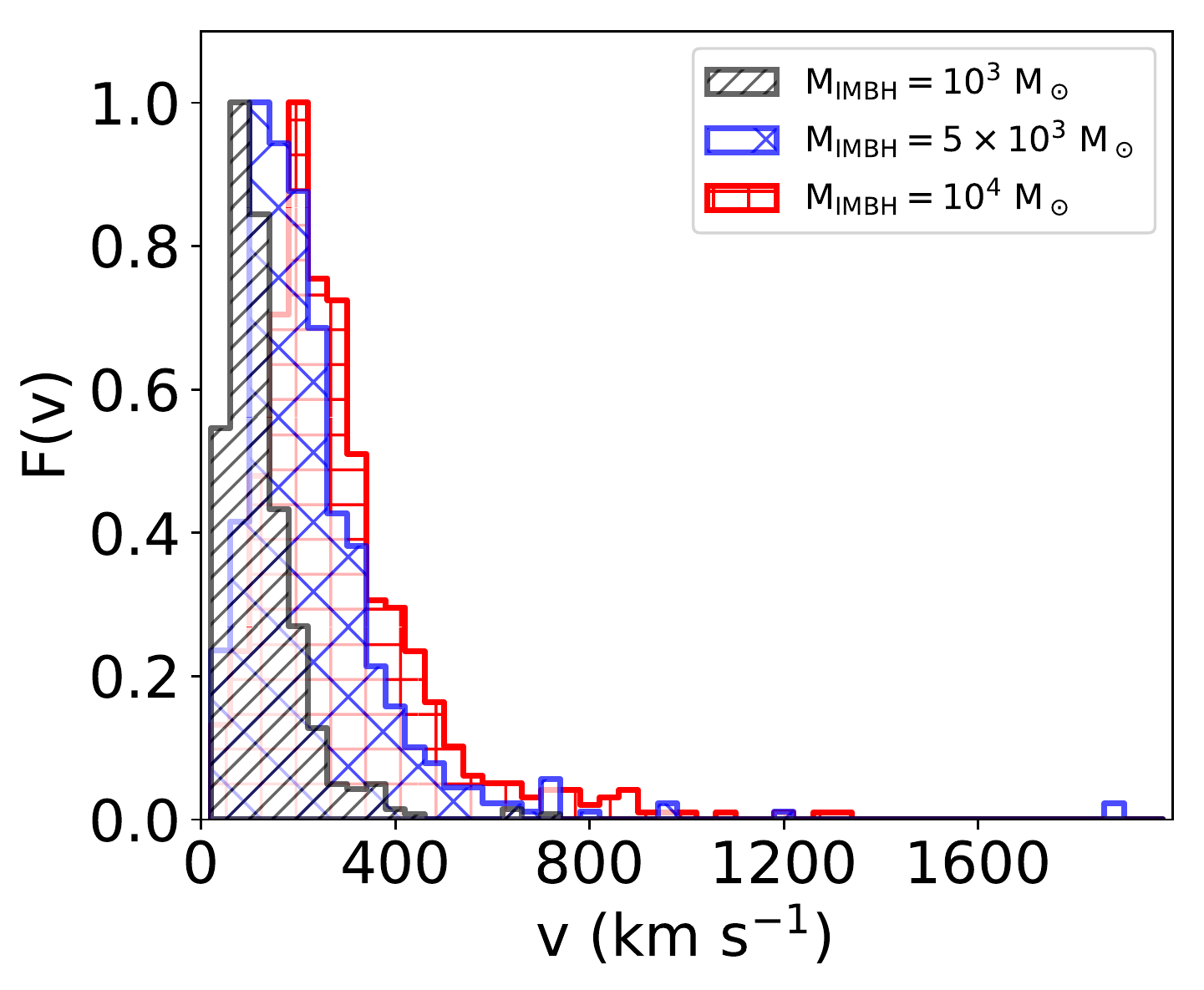}}
\subfloat{\includegraphics[scale=0.55]{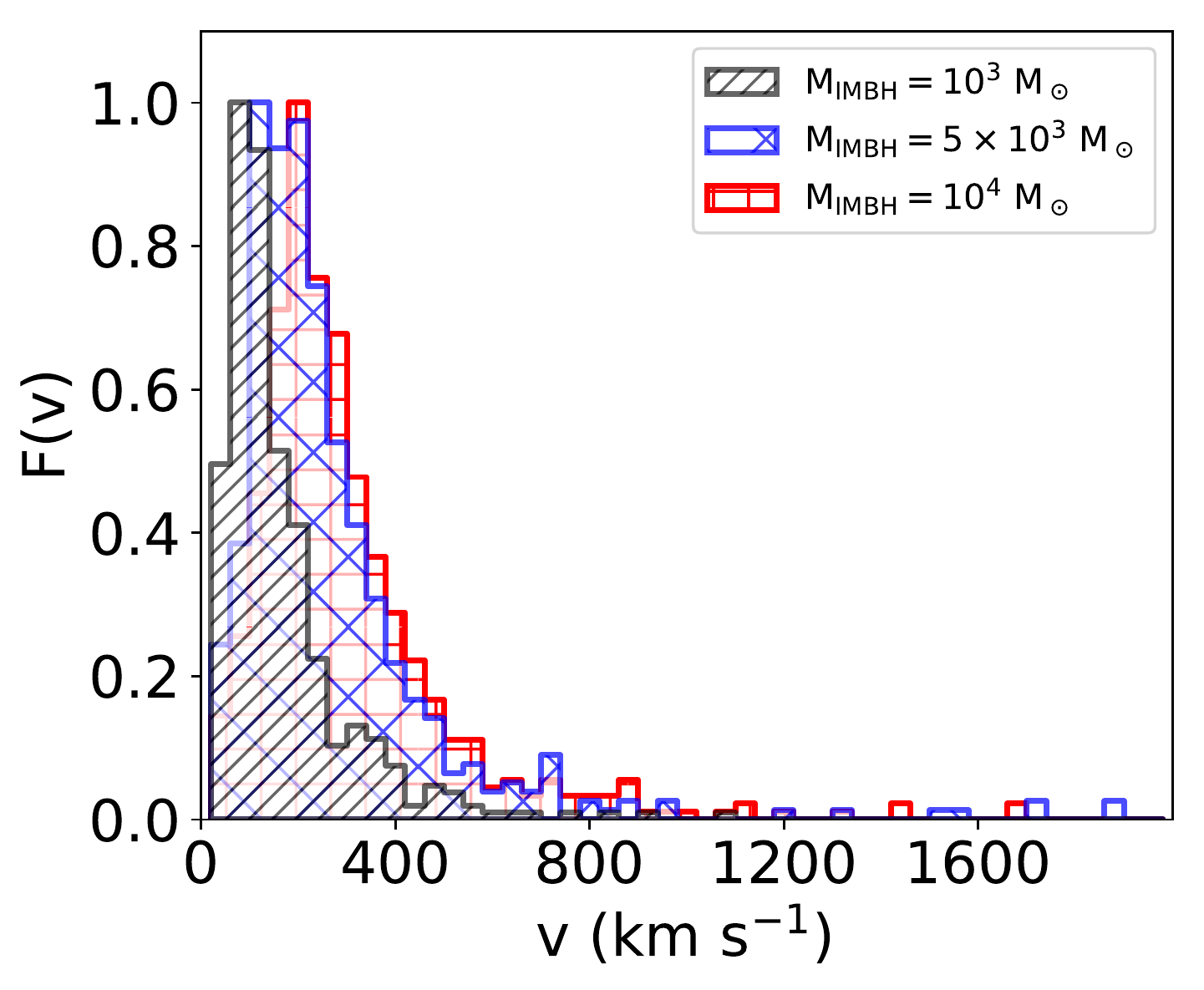}}
\end{minipage}
\caption{Velocity of ejected stars with $M_\mathrm{SMBH}=4\times 10^6$ M$_\odot$ (left) and $M_\mathrm{SMBH}=10^8$ M$_\odot$ (right), $a_{out}=0.01$ pc, $e_{out}=0.4$ and different IMBH masses. Only a few stars have velocities exceeding the local galactic escape speed (assuming a Milky-Way like galaxy).}
\label{fig:velocity}
\end{figure*}

The results of these numerical scattering experiments are shown in Figure~\ref{fig:tdeq12}. Each figure shows the distance of closest approach between the interloping single star (assumed to be massless relative to the SMBHs), in units of the tidal disruption radius for each SMBH. The solid lines show the simulated distributions for the primary SMBH, and the dashed lines sow the results for the secondary SMBH. As is clear from these figures, dynamical scattering results in disruption by the secondary SMBH a very small fraction of the time, less than $1$\% of the total number of random scatterings. This is the case independent of the binary mass ratio, semi-major axis and eccentricity. Thus, this supports our conclusion that the secular mechanism for the tidal disruption of stars by a lower-mass secondary SMBH considered in this paper is a considerably more efficient mechanism for probing SMBH-SMBH or SMBH-IMBH binarity in the most massive galaxies, relative to chaotic dynamical scattering of stars into the loss-cone of the secondary SMBH.

\subsection{Hypervelocity stars}

Stars with extreme radial velocities have recently been observed in the Galactic halo, the so-called hypervelocity stars (HVSs). First predicted by \citet{hills88} as a consequence of the tidal separation of binary stars by an SMBH, the fist HVS was observed by \citet{brw05} moving with a heliocentric radial velocity of $\sim 700\kms$. More than $20$ HVSs have now been confirmed in our Galaxy, with distances between $50$ and $120\kpc$ from the Galactic Centre and velocities up to $\approx 700\kms$ \citep{brw06,brw12,brw14,brw15}. Detailed calculations by \citet{yut03} showed that the ejection rate via the Hills mechanism is $\approx 10^{-6}-10^{-5}\yr^{-1}$ in the case of the Milky-Way.

While the ejection of HVSs due to the Hills mechanism still remains the favoured model for most of HVSs \citep{fgi18,fgu18,frs18}, alternative mechanisms have been proposed.  These include encounters with an SMBH binary \citep{yut03}, tidal interactions of star clusters with a single or binary SMBH \citep{cap15,fra16,fck17} and the dynamical evolution of a thin and eccentric disk orbiting the SMBH \citep{sub16}. HVSs are very rare but interesting objects. They can provide information both about their formation environment, the galactic potential in which they travel \citep{frl17}, ans also serve as proxies for planetary dynamics in extreme conditions \citep{frg17}. Upcoming data from \textit{Gaia} promise to shed definitively light on the origins of such objects in our Galaxy \citep{mar17}.

We have discussed that stars that get stripped from their orbit around the IMBH, which can have two fates. These stars can either be captured by the more massive SMBH and end up in direct orbit about it, or they can be ejected from the galactic nucleus. In the latter case, if their velocity exceeds the local escape speed, they become unbound with respect to the host galaxy and can travel far out through the galactic halo. Figure \ref{fig:velocity} shows the ejection velocities of stars scattered in the case $M_\mathrm{SMBH}=4\times 10^6$ M$_\odot$ (left) and $M_\mathrm{SMBH}=10^8$ M$_\odot$ (right), $a_{out}=0.01$ pc, $e_{out}=0.4$ and different IMBH masses. We find that in a Milky Way-like nucleus, the fraction of stars ejected with velocities larger than the local escape speed from the galaxy is $\approx 1\%$, a fraction that increases to $\approx 2\%$ in the case of $M_\mathrm{SMBH}=10^8$ M$_\odot$. Considering $N_*\approx 10$ stars in orbit around the IMBH, this leads to a rate of $\approx 10^{-8}-10^{-7}$ yr$^{-1}$, smaller than the \citet{hills88} rate. If the Milky-Way hosts an IMBH in its nucleus \citep{gua09}, some of the observed HVSs ($\approx 1$\%) may have originated due to the depletion of stars around the IMBH via perturbations from the primary SMBH.

\section{Discussions and Conclusions}
\label{sect:conc}

In this work, we study the fate of stars orbiting an IMBH secondary perturbed by a primary SMBH, as a function of the mass ratio of the SMBH-IMBH binary, the semimajor axis and the eccentricity of the IMBH orbit.  This is done by means of high-precision direct N-body simulations. We find that the presence of the secondary SMBH significantly alters the evolution of the stellar orbits.  For example, the region of parameter space corresponding to high star orbital inclinations with respect to the SMBH-IMBH orbital plane becomes depleted due to Kozai-Lidov oscillations initiating TDEs with the IMBH secondary. Consequently, the stars end up distributed in a torus-like shape with an extent that depends on the SMBH-IMBH mass ratio and the IMBH orbital parameters. This distribution may be resolved by instruments with an angular resolution on the order of the IMBH torus of stars. Such instruments could include the Keck telescope, Gemini and the Very Large Telescope \citep{lin15}.

We also show that the TDE rate can be as high as $\approx 10^{-4}-10^{-3}$ yr$^{-1}$, and that most TDEs occur on a timescale of $\approx 0.5$ Myr. The typical semimajor axis of stars plunging onto the IMBH depends on the details of the IMBH orbit and on the IMBH-SMBH mass ratio: the larger the mass ratio the larger the average semimajor axis of stars that undergo TDEs. 

We further argue that above a critical SMBH mass of $\approx 1.15 \times 10^8$ M$_{\odot}$ (in galaxies with bulges more massive than $\approx 4.15\times 10^{10}$ M$_{\odot}$), no TDE event can occur for typical stars in an old stellar population since the Schwarzschild radius exceeds the tidal disruption radius and any TDE will remain dark.  Consequently, no TDEs should be observed unless a lower-mass secondary SMBH or IMBH is also present, as studied in the present work.  In these massive galaxies, TDEs onto IMBHs can enlighten the galactic nucleus and potentially reveal the presence of a secondary SMBH/IMBH via the secular mechanism considered here. Finally, we have shown that the rate of ejected HVSs can be $\approx 1\%$ of the standard \citet{hills88} mechanism for a Milky-Way like nucleus, implying that some of the observed Galactic HVSs may have originated through the dynamical scenario presented in this work.

\section{Acknowledgements}

This research was partially supported by an ISF and an iCore grant. GF thanks Seppo Mikkola for helpful discussions on the use of the code \texttt{ARCHAIN}. Simulations were run on the \textit{Astric} cluster at the Hebrew University of Jerusalem.

\bibliographystyle{mn2e}
\bibliography{refs}

\end{document}